\documentclass[conference]{IEEEtran}
\IEEEoverridecommandlockouts

\usepackage{amsmath,amssymb,amsfonts}
\usepackage{algorithmic}
\usepackage{graphicx}
\usepackage{textcomp}
\usepackage{xcolor}
\usepackage{multirow} 
\def\BibTeX{{\rm B\kern-.05em{\sc i\kern-.025em b}\kern-.08em
    T\kern-.1667em\lower.7ex\hbox{E}\kern-.125emX}}
\begin{document}

\title{Development of a Mobile Application for at-Home Analysis of Retinal Fundus Images\\

}

\author{\IEEEauthorblockN{1\textsuperscript{st} Mattea Reid}
\IEEEauthorblockA{\textit{Division of Engineering Science} \\
\textit{University of Toronto}\\
Toronto, Canada \\
mattea.reid@mail.utoronto.ca}
\and
\IEEEauthorblockN{2\textsuperscript{nd} Zuhairah Zainal}
\IEEEauthorblockA{\textit{Computing and Information Systems} \\
\textit{Universiti Teknologi Brunei}\\
Bandar Seri Begawan, Brunei \\
B20220046@student.utb.edu.bn}
\and
\IEEEauthorblockN{3\textsuperscript{rd} Khaing Zin Than}
\IEEEauthorblockA{\textit{School of Information Technology} \\
\textit{King Mongkut’s University of Technology Thonburi}\\
Bangkok, Thailand \\
khaingzin.than@mail.kmutt.ac.th}
\and
\IEEEauthorblockN{4\textsuperscript{th} Danielle Chan}
\IEEEauthorblockA{\textit{Innovative Cognitive Computing Research Center} \\
\textit{Chinese International School of Hong Kong}\\
Hong Kong, China \\
daniellec2028@student.cis.edu.hk}
\and
\IEEEauthorblockN{5\textsuperscript{th} Jonathan H. Chan}
\IEEEauthorblockA{\textit{School of Information Technology} \\
\textit{King Mongkut’s University of Technology Thonburi}\\
Bangkok, Thailand \\
jonathan@sit.kmutt.ac.th}
}

\maketitle

\begin{abstract}
Machine learning is gaining significant attention as a diagnostic tool in medical imaging, particularly in the analysis of retinal fundus images. However, this approach is not yet clinically applicable, as it still depends on human validation from a professional. Therefore, we present the design for a mobile application that monitors metrics related to retinal fundus images correlating to age-related conditions. The purpose of this platform is to observe for a change in these metrics over time, offering early insights into potential ocular diseases without explicitly delivering diagnostics. Metrics analysed include vessel tortuosity, as well as signs of glaucoma, retinopathy and macular edema. To evaluate retinopathy grade and risk of macular edema, a model was trained on the Messidor dataset and compared to a similar model trained on the MAPLES-DR dataset. Information from the DeepSeeNet glaucoma detection model, as well as tortuosity calculations, is additionally incorporated to ultimately present a retinal fundus image monitoring platform. As a result, the mobile application permits monitoring of trends or changes in ocular metrics correlated to age-related conditions with regularly uploaded photographs.

\end{abstract}


\section{Introduction}
According to the National Eye Institute, 3.7 million Americans are projected to suffer from advanced age-related macular degeneration by 2030 [1]. Similarly, 4.3 million are projected to experience glaucoma, 11.3 million diabetic retinopathy, and 38.7 million cataracts. Since these medical conditions are increasingly treatable if detected early, there is an evident necessity for early diagnostic tools and procedures.

We therefore focus on developing a mobile application that facilitates the at-home capture and upload of retinal fundus images with the use of a small lens device. The mobile application provides the user with an overview of a variety of optical-health-related metrics associated with the uploaded fundus image. These metrics will be indicators that have been shown to correlate with several age-related conditions, which the user can monitor over a period of time to effectively supervise their eye health. The focus differs from similar platforms, as it encompasses surveilling long-term trends, rather than attempting to provide diagnostics from a single uploaded photograph.

Retinal fundus images are presently a tool able to identify a variety of medical conditions, such as diabetic retinopathy. Still, recent studies have increasingly shown their ability to detect early signs of age-related conditions. Early dementia and multiple sclerosis have been shown to be detectable from retinal fundus images through characteristics such as vessel tortuosity [2][3].

Retinal fundus images are generally captured professionally with the aid of large fundus cameras. Nevertheless, photos can be taken with a simple lens and a smartphone camera [4]. This is not a common practice, as it results in lower photo quality and does not eliminate the need for assistance from a medical professional. Although a mobile application is not a precise and reliable tool for diagnostics, it has potential utility in tracking trends from regularly taken images. Trends in metrics associated with age-related conditions could thereby be indicative of such conditions.
 \\

\section{Literature Review}

\subsection{Conditions Detectable Through Retinal Fundus Images}\label{AA}

Age-related macular degeneration (AMD) models have been developed to detect AMD with high accuracy, such as DeepSeeNet, which classifies fundus images with the AREDS Simplified Severity Scale, with an accuracy of 0.671 [5]. The DeepSeeNet model was compared to classifications performed by physicians, which received an accuracy score of 0.599. Beyond general classification, this model provides information on drusen size, the presence of pigmentary abnormalities, late AMD, geographic atrophy and central geographic atrophy [6].

Diabetic retinopathy (DR) retinal imaging can reveal DR through swelling of retinal veins or leakage of fluids or blood. Developments continue to be made on the current models to detect DR. One study aimed to improve on the DenseNet-121 CNN model using Bayesian approximation, achieving an accuracy of 0.9768 [7]. 

For glaucoma, retinal imaging can reveal damage to the optic nerve caused by intraocular pressure and thinning of the nerve fibre layer, both signs of the condition. Based on a 2025 review, models have been able to achieve a 96\% AUC in glaucoma diagnosis, and fundus images can provide a sensitivity of 92\% and specificity of 93\% [8]. 

Melanomas can also be detected through machine learning. A recent study conducted in 2024, aiming to detect uveal melanoma, achieved a Dice coefficient of 0.86 and a sensitivity of 1.00 [9].

Retinal imaging can additionally identify specific biomarkers associated with dementia, structural and vascular changes in the retina and alterations in nerve layers and blood vessels. Retinal thinning and amyloid beta deposits can also be indicative of dementia. This research is more recent and has achieved an AUC of 0.841 on test sets [2].

\subsection{Ocular Assessment Platforms}
Many mobile ocular assessment platforms have been developed for various diagnostic objectives. In 2016, Yin et al developed a cloud-based system to screen retinal fundus images and connect the resulting information to both the patient and ophthalmologist [10]. The platform focused on numerous diagnostics, with an emphasis on the connection to physicians and patients.  

Abdulhussein et al offer a review of 17 mobile eyecare applications from prior to 2022 [11]. These applications are split with 35\% focusing on glaucoma, 29\% on visual acuity, 24\% on age-related macular degeneration, and the remaining 12\% on a non-specified application. The applications are predominantly targeted toward the diagnostics of specific conditions, and vary in intended demographic.

A 2022 publication by Bernard et al outlines the use of a smartphone application to detect retinoblastoma through leukocoria [12]. The study collected 4356 photographs of pupils, and trained a model to detect leukocoria, ultimately yielding a sensitivity of 87\% and specificity of 73\%, among other analyses.

In 2023, Zang and Echegoyen developed a platform to detect dry eye disease (DED) [13]. The input taken is video footage of an individual’s blinking rates, and the resultant output is a 10-point score analysis for DED.

\subsection{Devices for At-Home Retinal Fundus Imaging}\label{AA}
A team from the University of California, Berkeley developed a mobile application focused on the detection of glaucoma, with the use of a 3D-printed clip-on lens [14]. The cost of the 3D-printed lens is roughly \$30 USD, underlining the lack of cruciality of expensive technology for this purpose. There exist numerous substitutes in terms of mobile retinal fundus image capture that are compatible with mobile phones, including the oDocs nun IR camera [15]. 
The platform described in this paper is designed to be compatible with any such at-home fundus-image-capturing device, as it allows for photo upload.

\section{Method}
For the development of this application, a model was trained to detect signs of macular edema and retinopathy from the Messidor dataset. The DeepSeeNet model was used to output information related to AMD. Tortuosity calculations were also performed.
\subsection{Retinopathy and Risk of Macular Edema}
A model was trained to detect the retinopathy grade and risk of macular edema with the Messidor data set.

\paragraph{Dataset} This consists of 1200 retinal fundus images, each labelled with a retinopathy score and a risk of macular edema [16]. The first 1100 were used for training and validation, with an 80\% and 20\% split. The remaining 100 for testing. The MAPLES-DR dataset was also used, a dataset containing 198 of the Messidor images and various features segmented for each of them [17].  A model trained on this dataset was used to compare with the results of the Messidor-based model. This was used to determine if the addition of the segmented features would result in better performance, despite the dataset being smaller. Testing was done with 58 images, and the remaining were split at 80\% for training and 20\% for validation. 

\paragraph{Model Architecture} For the Messidor model, a ResNet-50 backbone, pretrained on ImageNet, was used for feature extraction, then two classification heads for retinopathy grade and risk of macular edema.

The MAPLES-DR differs in that it uses EfficientNet-B0 for feature extraction as a base encoder and has a transformer backbone. Each segmented feature from the dataset has a learnable embedding. 

\paragraph{Data preprocessing} The images were first preprocessed with resizing to 224x224, and augmentation by flips, rotations, colour jitter and crops, as well as normalization. 

\paragraph{Training Procedure} Training was done for the first model using weighted cross-entropy losses summed across tasks, Adam optimizer (learning rate of 1e-4) with cosine annealing to adjust the learning rate over time, as well as early stopping. It additionally uses a frozen backbone for the first 5 epochs.

For the MAPLES-DR model, training was done with AdamW optimiser starting on non-encoded parameters for the first 3 epochs (learning rate of 2e-4), then the Efficient-Net encoder was unfrozen (learning rate of 1e-4 for encoder). Cosine annealing was also used.

\subsection{Signs of Age-Related Macular Degeneration}
A model to detect glaucoma was used from the DeepSeeNet [5][6][18], which has an accuracy score of 0.671, sensitivity of 0.590 and specificity of 0.930. Retinal specialists asked to provide the same diagnostics had an accuracy of 0.599, sensitivity of 0.512 and specificity of 0.916.

This model provides some binary metrics, such as the presence of pigmentary abnormalities, as well as some continuous gradings that can be tracked over time. This overall provides the user with an AMD prediction based on these symptoms analysed. This is useful as a monitoring tool, since it can provide data on worsening symptoms of AMD over time.

\subsection{Tortuosity Calculations}\label{SCM}
Tortuosity is a measure of the curvature of the vessels present in the photo, defined as the length of a vessel over a certain region divided by the Euclidean distance between points [19]. Increased vessel tortuosity has been shown to correlate with age-related conditions, such as dementia and multiple sclerosis.

To track this metric, first, when the retinal image is uploaded, it is run through a segmentation model to extract the vessels from the image [20]. After this, the vessels are skeletonised, to be one pixel in width and a tortuosity calculation is performed.

\section{Results}
\subsection{Retinopathy and Risk of Macular Edema}
The model trained on the Messidor dataset achieved a higher accuracy than the model trained with the MAPLES-DR dataset; therefore, in this case, the feature segmentation does not prove to be more useful than a more extensive database.  

\begin{table}[htbp]
\renewcommand{\arraystretch}{1.2}
\centering
\caption{Accuracy Scores Summary}
\label{tab:retino_accuracy}
\begin{tabular}{|c|c|c|c|}
\hline
\multirow{2}{*}{\textbf{Dataset}} & 
\multirow{2}{*}{\begin{tabular}[c]{@{}c@{}}\textbf{No. of Images Used} \\ \textbf{(Training and Validation)}\end{tabular}} & 
\multicolumn{2}{c|}{\textbf{Accuracy}} \\ \cline{3-4}
 & & \textbf{\textit{Retinopathy}} & \begin{tabular}[c]{@{}c@{}}\textbf{\textit{Macular}} \\ \textbf{\textit{Edema}}\end{tabular} \\ \hline
Messidor  & 1100 & 0.60 & 0.86 \\ \hline
Messidor  & 700  & 0.55 & 0.58 \\ \hline
MAPLES-DR & 138  & 0.36 & 0.72 \\ \hline
\end{tabular}
\end{table}

Table 1 shows the accuracy scores for the test images of the Messidor model trained on 1100 images, versus 700, both compared to the MAPLES-DR model. Overall, the accuracy of the Macular Edema detection is higher for the MAPLES-DR model than the Messidor trained on 700 images. 

Figures 1 and 2 represent confusion matrices for the model trained on 1100 of the Messidor images. It demonstrates the imbalance in the dataset, which additionally impacts the performance of the models.

\begin{figure}[htbp]
\centerline{\includegraphics[width= 0.45\textwidth]{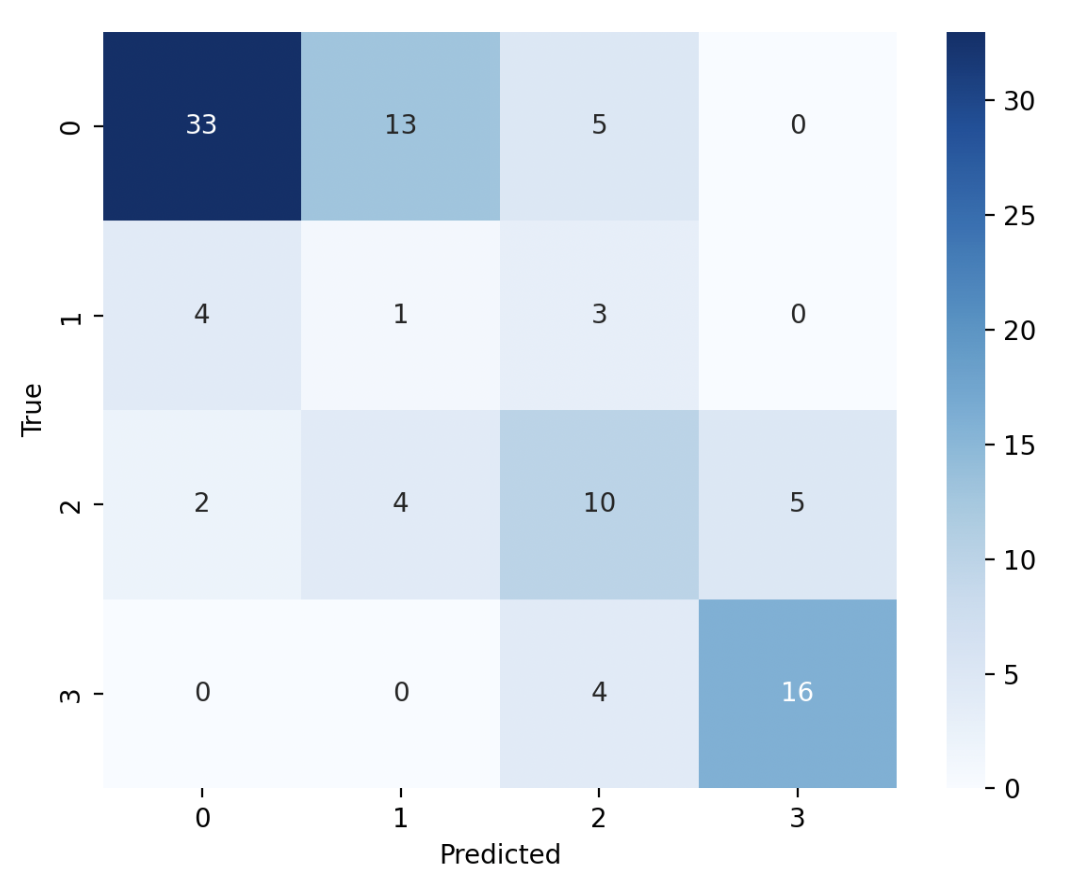}}
\caption{\centering A confusion matrix for the retinopathy grade predicting model, trained on 1100 images and tested on 100 images from the Messidor dataset. The different grades are represented by the numbers 0-3, with 0 indicating no retinopathy, and 3 indicating severe retinopathy.}
\label{fig}
\end{figure}

\begin{figure}[htbp]
\centerline{\includegraphics[width= 0.45\textwidth]{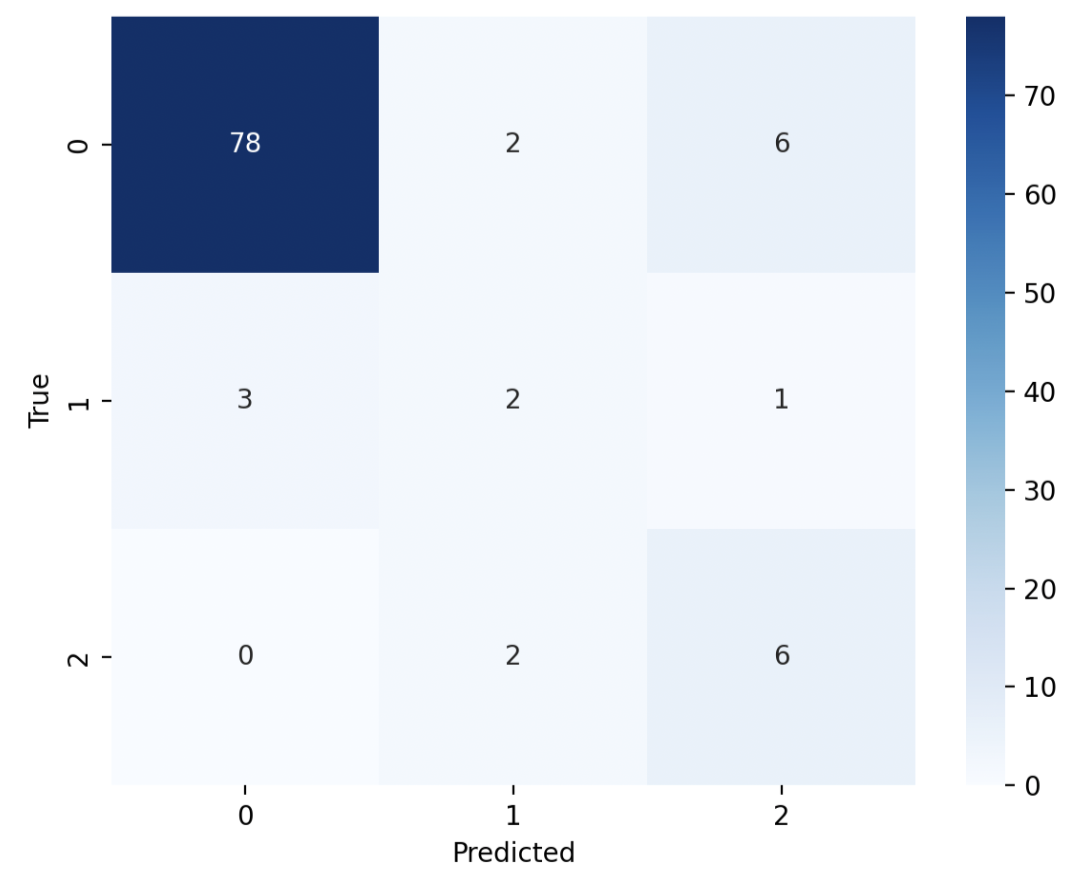}}
\caption{\centering A confusion matrix for the risk of macular edema predicting model, trained on 1100 images and tested on 100 images from the Messidor dataset. The numbers 0-2 represent the different risks of edema, with 0 being the lowest and 2 the highest.}
\label{fig}
\end{figure}

\section{App Design}

The following is a description of various crucial features of the mobile application created. First, at-home retinal image capture was enabled for users to take or upload retinal photos directly from their smartphone camera, providing compatibility with any at-home retinal fundus image-capturing device. Integrated models then analyze each image and extract key health-related metrics based on public training datasets, with the following metrics presented to the user when a photo is uploaded:
\begin{itemize}
    \item Average tortuosity of vessels
    \item Maximum tortuosity
    \item Risk of macular edema
    \item Retinopathy grade
    \item Age-Related Macular Degeneration Grade
    \begin{itemize}
        \item Drusen Score
        \item Pigmentary abnormalities
        \item Geographic atrophy
        \item Central Geographic Atrophy
    \end{itemize}
\end{itemize}

The glaucoma score is 1 if glaucoma is detected, and 0 otherwise. The retinopathy grade varies from 0-3, and the risk of macular edema from 0-2, with higher numbers indicating increased severity. Tortuosity does not have discrete levels; it is rather on a continuous scale. However, these are presented to the user as levels rather than numbers to increase clarity and ease of use. In addition, information is given about drusen score, pigmentary abnormalities, the presence of geographic atrophy, and the presence of central geographic atrophy. The application then incorporates a generative AI interpretation, which calls DeepSeek to aid users with the interpretation of results, as well as provide suggestions for future action. Figure 3 provides the prompt given to DeepSeek to aid in the interpretation of metrics output by models.

\begin{figure}[htbp]
\centerline{\includegraphics[width= 0.45\textwidth]{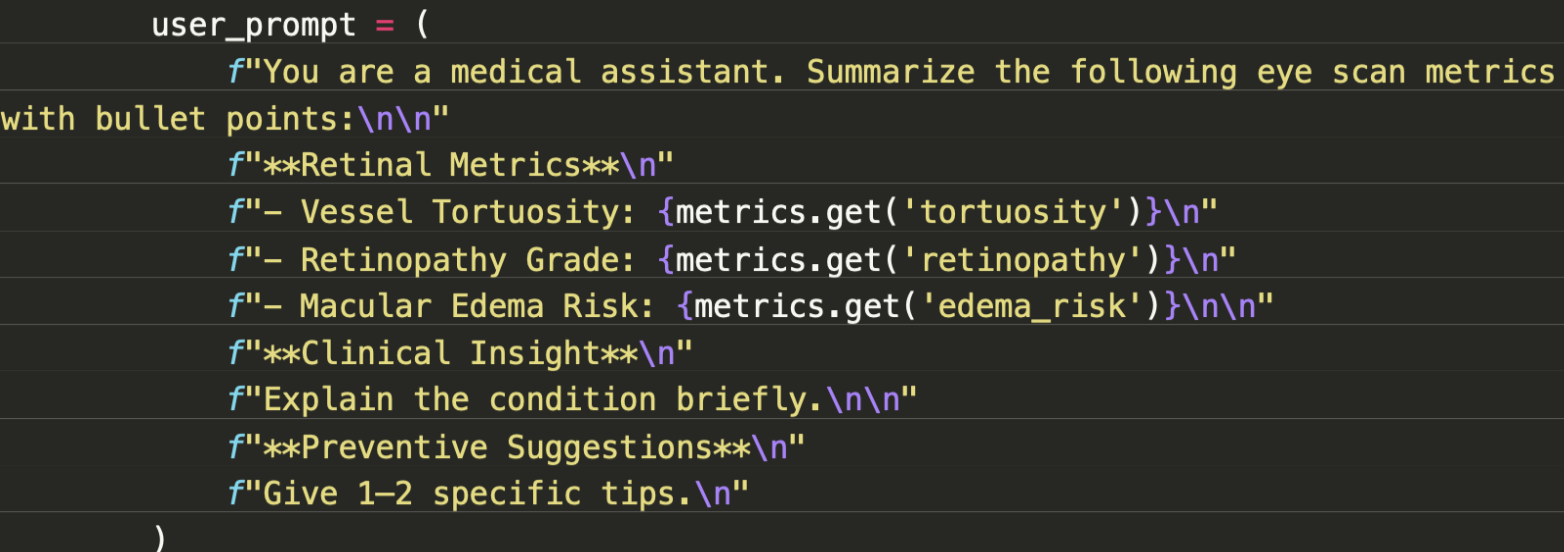}}
\caption{\centering An excerpt of the application program highlighting the prompt given to DeepSeek regarding three metrics output by various models. The prompt requests a bullet point summary of the interpretation of the metrics, followed by a brief summary of the pertinent medical conditions, as well as suggested courses of action. Note that although not visible in this figure, the platform provides warnings to the user that the generative AI feature does not represent a medical professional and should not be treated as such. 
}
\label{fig}
\end{figure}

The visual display of the metrics aims to facilitate comprehension of results, with visual indicators such as colour-coded severity levels and simple explanations of each metric. There is additional emphasis on change over time, with features highlighting abnormal shifts in various metrics. As this application is targeted towards older adults, there is a focus on ensuring ease of use, highlighted by tutorial pages that provide insight into the functionality of the app. Information pages also provide insights into the signs of age-related conditions and offer guidance on how to interpret the associated results. Figure 4 illustrates the visual diagrams related to five different eye diseases, providing the user with a concise, easily interpretable description of the age-related diseases detectable through retinal imaging.

\begin{figure}[htbp]
\centerline{\includegraphics[width= 0.45\textwidth]{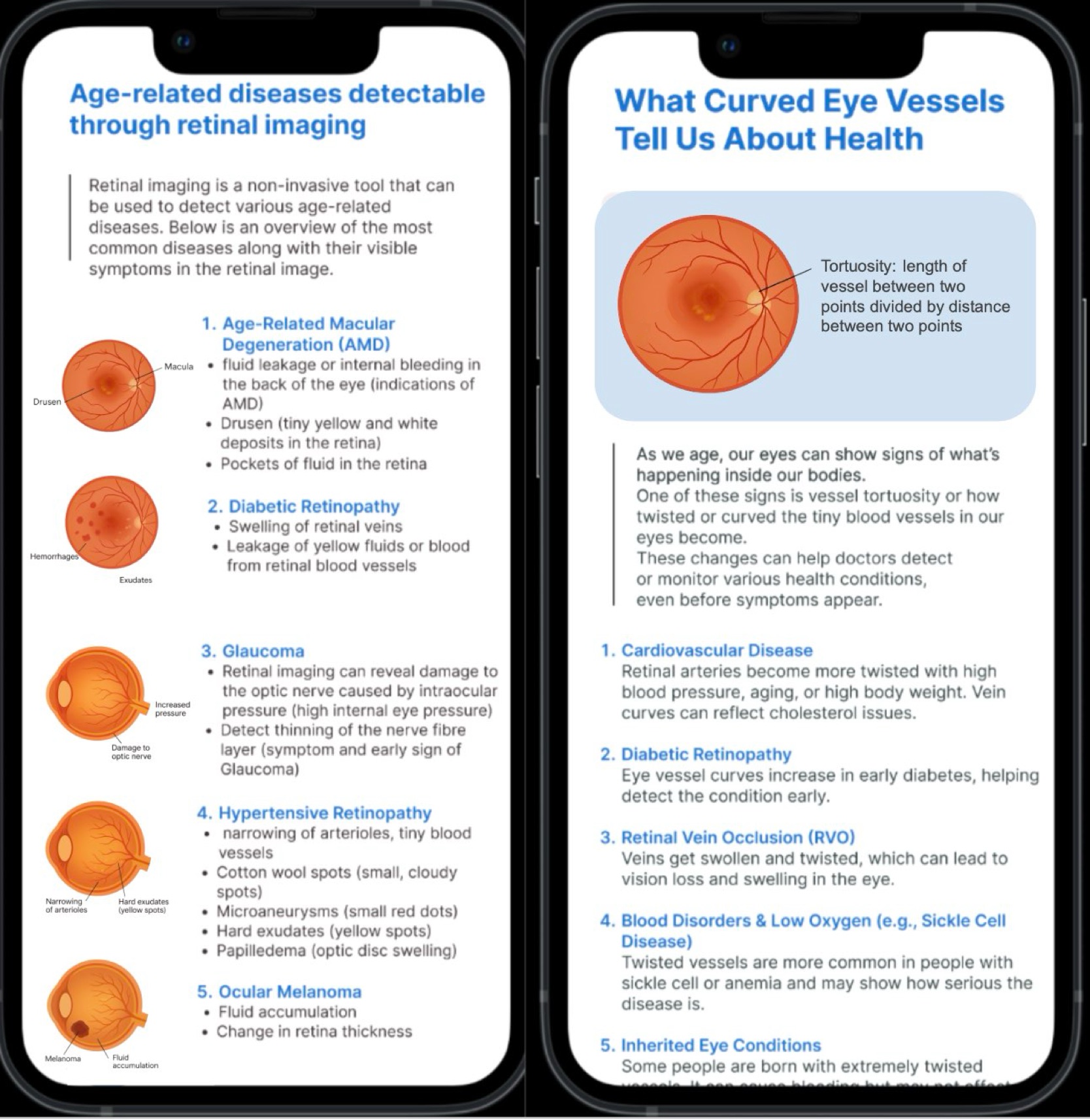}}
\caption{\centering Sample screenshots of the user interface of our application describing age-related eye diseases with diagrams [21], in addition to a sample information page detailing the interpretation of curved eye vessels on eye-health.}
\end{figure}
Calendar and timeline views enable progress tracking to present trends and patterns that help users monitor potential signs of age-related conditions (e.g., early dementia, multiple sclerosis). 

Users can additionally add notes after each scan, export reports for personal or medical use and share data securely with healthcare providers if desired. Each user has a personalized profile with scan history that uses Firebase Auth for secure authentication and data privacy. Offline and cloud support allow offline image capture, with upload queued for later. Firebase Storage ensures encrypted, reliable cloud data storage.

\section{Discussion}
This paper provides an overview of a mobile application meant to enable tracking of metrics associated with ocular health. Since artificial intelligence models cannot work independently from medical professionals, this platform is meant to track the progression of retinal characteristics rather than attempt to provide diagnostics. Potential uses include personal records and monitoring of relative eye health, in addition to conducting further studies on the progression of eye disease.

The accuracy of the retinopathy and risk of macular edema detection models is 0.60 and 0.86, respectively, partially due to the imbalance in the dataset, which is underlined in Figures 1 and 2. The use of the MAPLES-DR dataset ultimately did not provide a higher accuracy than simply training on a larger dataset. Nevertheless, the macular edema detection of the MAPLES-DR model has a higher accuracy than that of the Messidor model trained on 700 images. It can thus be concluded that the feature segmentation is beneficial in the reduction of the number of images needed for training. Despite this, the runtime is not reduced significantly as the model requires the analysis of additional feature images.

Limitations of this platform are related to the performance of the models, which have the opportunity for further improvement. This process may include training on further data. There is additional potential for the analysis of supplemental ocular characteristics and metrics. Since research regarding retinal fundus images and age-related conditions is continually progressing, this application may be updated accordingly. The scope could also be increased to screen for non-age-related conditions, as well as implementing a continuous grading to further distinguish trends on the calendar feature. This, although not representative of the grading levels for medical conditions, could allow for a more insightful tracking process.

Applications of this software other than personal use include providing a monitoring system compatible with healthcare professionals to monitor the eye health of patients, in order to increase accessibility for individuals with mobility difficulties. A second potential utilisation is to provide a platform for research regarding the progression of retinal characteristics with the progression of age-related conditions. The accessibility and ease-of-use of the application can facilitate regular photo capture and upload for such further research.

\section*{Conclusion}
The purpose of the outlined mobile application is to facilitate the tracking and monitoring of metrics extracted from retinal fundus images, with the hopes of detecting trends that correlate to age-related conditions. Features of this application include a photo upload, models that analyse and output characteristics of images, a generative AI response to help interpret results, as well as a calendar feature to track these metrics over time. Potential applications for this software include personal use, use from medical professionals, as well as providing a platform for research.

\section*{Acknowledgment}

The authors would like to thank the Division of Engineering Science at the University of Toronto, the School of Information Technology at King Mongkut’s University of Technology, as well as Mitacs for providing funding for this project. This project was completed at the Innovative Computing Computing (IC2) lab at  King Mongkut’s University of Technology Thonburi, with the assistance of Almas Toh-i and Ngwe Yee Pearl Ou.

\vspace{12pt}


\begin{thebibliography}{00}
\bibitem{b1} National Eye Institute, Eye Disease Statistics Fact Sheet, National Institutes of Health, Washington, DC, Fact Sheet 2014, Mar. 2014. [Online]. Available:https://www.nei.nih.gov/sites/default/files/
2019-04/NEI\_Eye\_Disease\_Statistics\_Factsheet\_2014\_V10.pdf [Accessed: Sept 14, 2025].

\bibitem{b2}C. D. DeBuc, “Retinal imaging to detect Alzheimer disease: 
Machine learning model,” Ophthalmology Times Europe. [Online].Available:https://europe.ophthalmologytimes.com/view/retinal-imaging-to-detect-alzheimer-disease-machine-learning-model[Accessed: May 29, 2025].

\bibitem{b3} Y. Jiang, Q. Chen, D. Shi, S. Miao, Y. Liu, J. Wang, L. Liu, Y. 
Chen, and R. Wang, “Association of retinal microvascular curve tortuosity and multiple sclerosis: A cross-sectional analysis from the UK Biobank,” NeuroImage: Clinical, vol. 88, p. 105753, 2024. [Online]. Available: https://pubmed.ncbi.nlm.nih.gov
/38996710/.

\bibitem{b4} American Academy of Ophthalmology, “How to take retinal 
images with a smartphone,” American Academy of Ophthalmology (AAO), Aug. 2, 2022. [Online]. Available: https://www.aao.org/eye-health/tips-prevention/how-to-take-retinal-images-smartphone. [Accessed: May 29, 2025].
\bibitem{b5} Y. Peng, S. Dharssi, Q. Chen, T. Keenan, E. Agron, W. Wong, E. 
Chew, and Z. Lu, “DeepSeeNet: A deep learning model for automated classification of patient based age-related macular degeneration severity from color fundus photographs,” Ophthalmology, vol. 126, no. 4, pp. 565–575, Apr. 2019.

\bibitem{b6} NCBI NLP, “DeepSeeNet: A deep learning framework for 
classifying patient-based age-related macular degeneration severity in retinal color fundus photographs,” GitHub. [Online]. Available: https://github.com/ncbi-nlp/DeepSeeNet. [Accessed: May 29, 2025].

\bibitem{b7}M. Akram, M. Adnan, S. F. Ali, et al., “Uncertainty-aware diabetic 
retinopathy detection using deep learning enhanced by Bayesian approaches,” Sci. Rep., vol. 15, no. 1342, 2025, doi: 10.1038/s41598-024-84478-x.

\bibitem{b8}X. C. Ling, H. S.-L. Chen, P.-H. Yeh, Y.-C. Cheng, C.-Y. Huang, 
S.-C. Shen, and Y.-S. Lee, “Deep learning in glaucoma detection and progression prediction: A systematic review and meta-analysis,” Biomedicines, vol. 13, no. 2, p. 420, 2025, doi: 10.3390/biomedicines13020420.
\bibitem{b9} J. Ma, S. P. Iddir, S. Ganesh, D. Yi, and M. J. Heiferman, 
“Automated segmentation for early detection of uveal melanoma,” Can. J. Ophthalmol., vol. 59, no. 6, pp. e784–e791, 2024, doi: 10.1016/j.jcjo.2024.04.003.
\bibitem{b10} F. Yin, B.-H. Lee, A. Yow, Y. Quan, and D. Wong, “Automatic ocular disease screening and monitoring using a hybrid cloud system,” in Proc. IEEE Int. Conf. Internet of Things (iThings), Green Computing and Communications (GreenCom), Cyber, Physical and Social Computing (CPSCom), Smart Data (SmartData), 2016, pp. 263–268. doi: 10.1109/iThings-GreenCom-CPSCom-SmartData.2016.6.
\bibitem{b11} D. Abdulhussein, M. Abdul Hussein, M. Szymanka, and S. Farag, “A systematic review of the current availability of mobile applications in eyecare practices,” Eur. J. Ophthalmol., vol. 33, no. 2, pp. 754–766, 2022, doi: 10.1177/11206721221131397.
\bibitem{b12} A. Bernard, S. Z. Xia, S. Saleh, T. Ndukwe, J. Meyer, E. Soloway, M. Sintayehu, B. T. Ramet, B. Tadegegne, C. Nelson, and H. Demirci, “EyeScreen: Development and potential of a novel machine learning application to detect leukocoria,” Ophthalmol. Sci., vol. 2, no. 3, p. 100158, 2022, doi: 10.1016/j.xops.2022.100158.
\bibitem{b13} S. Zhang and J. Echegoyen, “Design and usability study of a point of care mHealth app for early dry eye screening and detection,” J. Clin. Med., vol. 12, no. 20, p. 6479, 2023, doi: 10.3390/jcm12206479.
\bibitem{b14} K. Abedrabbo, Lumos Lens Models: Test1.0.stl, GitHub. 
[Online]. Available: https://github.com/karimabedrabbo/Lumos [Accessed: Jul. 2, 2025].
\bibitem{b15} F. Hafiz, R. J. Chalakkal, S. C. Hong, G. Linde, R. Hu, B. 
O’Keeffe, and Y. Boobin, “A new approach to non-mydriatic portable fundus imaging,” Expert Rev. Med. Devices, vol. 19, no. 4, pp. 303–314, Apr. 2022, doi: 10.1080/17434440.2022.2070004.
\bibitem{b16} J. Decencière et al., “Feedback on a publicly distributed 
database: The Messidor database,” Image Anal. Stereol., vol. 33, no. 3, pp. 231–234, Aug. 2014, doi: 10.5566/ias.1155.
\bibitem{b17} G. Lepetit-Aimon, C. Playout, M.-C. Boucher, et al., 
“MAPLES-DR: MESSIDOR anatomical and pathological labels explainable screening of diabetic retinopathy,” Sci. Data, vol. 11, p. 914, 2024, doi: 10.1038/s41597-024-03739-6.
\bibitem{b18} T. Keenan, S. Dharssi, Y. Peng, Q. Chen, E. Agron, W. Wong, Z. 
Lu, and E. Chew, “A deep learning approach for automated detection of geographic atrophy from color fundus photographs,” Ophthalmology, 2019.
\bibitem{b19} V. Kashyap, R. Gharleghi, D. D. Li, et al., “Accuracy of vascular tortuosity measures using computational modelling,” Scientific Reports, vol. 12, no. 865, 2022, doi: 10.1038/s41598-022-04796-w.
\bibitem{b20} A. Galdran, A. Anjos, J. Dolz, et al., “State-of-the-art retinal vessel segmentation with minimalistic models,” Sci. Rep., vol. 12, no. 6174, 2022, doi: 10.1038/s41598-022-09675-y.
\bibitem{b21}OpenAI, \textit{AI-generated medical illustrations of retinal diseases (Age-Related Macular Degeneration, Diabetic Retinopathy, Glaucoma, Hypertensive Retinopathy, Ocular Melanoma)}, created with ChatGPT (GPT-5 with DALL·E), Sept. 2025.


\end{thebibliography}
\end{document}